\documentclass[useAMS,usenatbib]{mn2e}

\usepackage{amsmath}
\usepackage{amssymb}
\usepackage[dvipsnames]{xcolor}

\usepackage{graphicx}
\usepackage{graphics}
\usepackage{hyperref}
\hypersetup{
     colorlinks   = true,
     citecolor    = blue
}

\def\apj{{ ApJ}}
\def\apjl{{ ApJL}}
\def\apjs{{ ApJS}}

\def\aap{{ A\&A}}
\def\aj{{ AJ}}
\def\mnras{{ MNRAS}}
\def\araa{{ ARA\&A}}

\def\nat {{ Nature}}

\def\physrep{{ Physics Reports}}

\def\prd{{ Phys. Rev. D}}

\def\procspie{{Proc. SPIE}}
\def\nar{{New Astronomy Reviews}}

\def\mt{\mathrm}
\def\d{{\mathrm{d}}}
\def\tp{\widetilde{P}}
\newcommand{\myemail}{wenbinlu@astro.as.utexas.edu}

\title[stellar disruptions and the black hole event horizon]{Stellar
  disruption events support the existence of the black hole event
  horizon} 
\author[Lu, Kumar, \& Narayan]
  {Wenbin Lu$^1$\thanks{\myemail},
  Pawan Kumar$^1$\thanks{pk@astro.as.utexas.edu},
  Ramesh Narayan$^2$\thanks{rnarayan@cfa.harvard.edu}\\
  $^1$Department of Astronomy, University of Texas at Austin, Austin,
TX 78712, USA\\
  $^2$Harvard-Smithsonian Center for Astrophysics, 60 Garden Street,
  Cambridge, MA 02138, USA}

\pagerange{\pageref{firstpage}--\pageref{lastpage}} \pubyear{2015}
\def\LaTeX{L\kern-.36em\raise.3ex\hbox{a}\kern-.15em
    T\kern-.1667em\lower.7ex\hbox{E}\kern-.125emX}

\begin{document}
\label{firstpage}
\maketitle

\begin{abstract}
Many black hole (BH) candidates have been discovered in X-ray
binaries and in the nuclei of galaxies. The prediction of Einstein's
general relativity is that BHs have an event horizon
--- a one-way membrane through which particles fall into the BH but
cannot exit. However, except for the very few nearby supermassive BH
candidates, our telescopes are unable to resolve and 
provide a direct proof of the event horizon. Here, we 
propose a novel observation that supports the existence
of event horizons around supermassive BH candidates 
heavier than $10^{7.5}M_{\rm \odot}$. Instead of
an event horizon, if the BH candidate has a hard surface, when a star
falls onto the surface, the shocked baryonic gas will form a radiation 
pressure supported envelope that shines at the Eddington
luminosity for an extended period of time from months to years. We
show that such emission has already been ruled out by 
the Pan-STARRS1 3$\pi$ survey if supermassive BH 
candidates have a hard surface at radius larger than $(1 +10^{-4.4})$
times the Schwarzschild radius. Future observations by LSST should be
able to improve the limit to $1 +10^{-6}$.
\end{abstract}


\section{Introduction}\label{sec:intro}
A black hole (BH) forms when no other force can uphold gravity and
everything collapses down to a point of roughly Planck size $\sim 
10^{-33}\rm\ cm$. The prediction of Einstein's
general relativity is that the point mass must be enclosed inside an
event horizon, through which matter, energy and light can enter from
outside, but nothing can exit. For a non-spinning BH, the size of the event
horizon, or the Schwarzschild radius, is
proportional to the mass $M = 10^7 M_7M_{\rm \odot} $ as 
\begin{equation}
  \label{eq:2}
      r_{\rm S} \equiv \frac{2GM}{c^2} = 3.0 \times10^{12} M_7 \rm\ cm.
\end{equation}
Over the past 30 years, BH candidates have been found and classified 
according to their masses, with stellar-mass candidates having a few up to
tens of $M_{\rm \odot}$ and supermassive candidates with masses
$\sim10^6$-$10^{10}M_{\rm \odot}$. Proving the existence of the defining 
characteristic of BHs  --- the event horizon --- would provide crucial
support for Einstein's general relativity. However, BH event horizons
are usually too small for our telescopes to resolve. 

Nearly all galaxies have a central massive object (CMO) of
mass $\sim10^6$-$10^{10}M_{\rm \odot}$
\citep[e.g.][]{2013ARA&A..51..511K}. The nature of the CMOs is
important in many 
astrophysical fields, e.g. active galactic nuclei, galaxy evolution,
gravitational wave, etc. CMOs are widely {\it believed} to be BHs, due to the
following reasons \citep[many of which have been discussed by][]{
2008NewAR..51..733N}.

(1) If the mass of a compact object exceeds the maximum neutron star
mass $M_{\rm NS,max}\sim3M_{\odot}$, there is no known force that can hold
it up from collapsing.
(2) Since active galactic nuclei are powered by
accretion (or gravitational potential energy), the central
mass-gaining object or cluster is expected to undergo collapse
and eventually turn into a BH, if there is no exotic force supporting
gravity \citep[e.g.][]{1984ARA&A..22..471R}. 
(3) In the absence of an event horizon, the kinetic energy of the
infalling gas will be converted into radiation
inside or on the surface of the CMO. For the two
nearby CMOs at the centers of the Milky Way and M87 (Sgr A* and 
M87*), if they do
not have event horizons and are in thermal dynamic equilibrium, this
amount of radiation ($\sim\dot{M}c^2$) overproduces the observed 
infrared flux by a factor of 10--100 \citep{2009ApJ...701.1357B,
  2015ApJ...805..179B}. (4) The Event Horizon Telescope (EHT) images
of Sgr A* and M87* in the millimeter wavelength so far are consistent
with a point source of radius $\lesssim2$-$2.5r_{\rm S}$
\citep{2008Natur.455...78D, 2012Sci...338..355D}, which roughly
corresponds to the apparent size of the photon capture radius
(``BH shadow''), so a hard surface at radius
significantly larger than $1.5r_{\rm S}$ has been ruled out
\citep{2006ApJ...638L..21B}. As the sensitivity  
and resolution of EHT improve, future images will be compared to
realistic accretion flow models and will directly test the spacetime
metric. (5) The LIGO detections of gravitational wave bursts
(e.g. GW150914) are consistent with merging stellar-mass BHs
\citep{2016PhRvL.116v1101A}.

While the reasons above are certainly strong, one may argue: reason (1)
may not apply to CMOs because their compactness is unknown and there
might be some mechanism/material that can support them from
collapsing; reason (2) does not rule out many classes of BH
alternative models either \citep[e.g. boson stars and gravastars,
][]{2003CQGra..20R.301S, 2004PNAS..101.9545M}; reasons (3) and (4)
apply to only a few nearby CMOs (due to our telescopes' finite resolution).
As for reason (5), gravitational wave generation calculations for
binary merger events for alternative models of BHs (objects without
event horizon) have only been done in the ringdown phase where the
differences only appear in the late-time secondary pulses in the
high-compactness limit \citep[e.g.][]{2016PhRvD..94h4002Y}. Future
work on the gravitational wave emission
during the plunge and merger phase, combined with higher
signal-to-noise ratio LIGO detections, will put better constraints on
the alternative models. 

In this paper, we propose a novel observation that places
stringent constraints on the possible location of a hard surface
around CMOs and hence strongly argues for them being BHs with event horizons.

\section{The Idea}
Stars can be driven into nearly radial orbits towards
the CMO by different processes, e.g. two-body relaxation, resonant
relaxation, massive perturbers, non-spherical
potential \citep{2005PhR...419...65A}. If the CMO is compact enough,
stars could reach down to a critical radius 
where the tidal gravity exceeds the star's self-gravity, causing a
tidal disruption event (TDE). The Newtonian tidal disruption 
radius is given by \citep[e.g.][]{1988Natur.333..523R}
\begin{equation}
  \label{eq:1}
      \frac{r_{\rm T}}{r_{\rm S}} \simeq 5.0  r_{\rm *} m_{\rm
      *}^{-1/3} M_7^{-2/3},
\end{equation}
where the star's mass and radius are expressed in units of solar mass
and solar radius, $M_{\rm *} = m_{\rm *} M_{\rm  \odot}$ and
$R_{\rm *} = r_{\rm *} R_{\rm \odot}$. When $M\gtrsim10^{7.5}M_{\rm \odot}$,
the Newtonian tidal disruption radius is not applicable and a full
general relativistic treatment is necessary
\citep[e.g.][]{2012PhRvD..85b4037K, 2016arXiv161103036S}.
When the star crosses $r_{\rm T}$, the tidal gravity of the CMO causes
a spread of specific orbital energy across the star, which leaves
roughly half of the star in bound orbits and the other half
unbound. Then the fall-back gas forms a thick accretion disk
which produces months-long optical/UV luminosity
$10^{44}$-$10^{45}\rm\ erg/s$ 
observable at cosmological distances. Recently, a few dozen of such
TDE candidates have been observed from various surveys carried out
in the optical, UV and soft X-ray wavelengths, giving a TDE rate of
$\sim 10^{-5}\rm\ yr^{-1}\ galaxy^{-1}$, consistent with but somewhat lower
than theoretical estimates \citep[see the review
by][]{2015JHEAp...7..148K}.

The star in a TDE can be used as a test particle to probe the
nature of the CMO, because it reaches very close to $r_{\rm S}$. In
this paper, we consider the observational consequences 
of star-CMO close encounters if the CMO does not possess an event
horizon. We assume the CMO's radius to be
\begin{equation}
  \label{eq:9}
  r_0=\eta r_{\rm S},
\end{equation}
where $\eta>1$ is a free parameter. If the CMO has a hard surface,
then $r_0$ is the surface radius; if the CMO is a diffuse cluster of
non-luminous particles or objects, then $r_0$ is defined as the
half-mass radius and $M$ is the mass enclosed within radius
$r_0$. There are two possibilities in the hard-surface scenario (and both
are considered in this paper): if
the CMO is made of ordinary matter, the Buchdahl limit gives
$\eta>9/8$ \citep{Buchdahl59}; if exotic forces are allowed,
$\eta$ can be extremely close to 1.

From eq. (\ref{eq:1}), we know that TDEs with accretion disk formation
are only possible if the radius of the CMO is smaller than $r_{\rm T}$, i.e.
\begin{equation}
  \label{eq:25}
  \eta M_7^{2/3} < 5.0  r_{\rm *} m_{\rm *}^{-1/3}.
\end{equation}
Therefore, considering the fact that TDEs from CMOs of roughly
$10^6M_{\rm \odot}$ have been observed, we obtain an upper
limit $\eta < 30$. The nature
of CMOs should not depend on their 
masses, so we only consider the parameter space\footnote{A small
  fraction of CMOs could be of heterogeneous nature, but they are not
  the focus of this paper.} $1<\eta < 30$. The upper limit on $\eta$ can rule out
CMOs being clusters of brown dwarfs or stellar remnants (white
dwarfs, neutron stars and stellar-mass BHs), because the
lifetime due to collisions or evaporation is much shorter than $10\rm\
Gyr$ \citep{1998ApJ...494L.181M}. Other alternative models cannot be
ruled out yet, such as objects with a hard surface supported by exotic
forces \citep[e.g. gravastars,][]{2004PNAS..101.9545M} or the
configuration from collapse of self-interacting scalar fields 
\citep[boson stars, see the review
  by][]{2003CQGra..20R.301S} or clusters of very low mass BHs
  \citep[$\lesssim 10^{-6} M_7^2 (\eta/30)^{3/2} M_{\rm \odot}$ if we
  use the Newtonian evaporation rate of][]{1998ApJ...494L.181M}. 

In the absence of an event horizon, when a TDE occurs, the kinetic
energy of the baryons accreted onto the CMO is converted to thermal
energy which should be radiated away over a certain period of
time. Regardless of the nature\footnote{We 
  assume that the baryonic gas is incorporated into the CMO's 
pre-existing exotic material slowly enough that shocks can form, and
that the shocked gas will expand due to its own pressure gradient.} of
the CMO, the 
accreted gas will be shocked when colliding with itself or the
possible hard surface. Then the shocked gas will form a hot envelope
surrounding the CMO. As far as we know from baryonic physics, the 
layer of stellar debris must  
be supported by radiation pressure. One could infer the (non-)existence
of this radiating stellar debris layer by multi-wavelength observations of
TDEs. Disproving the existence of such emission supports
the existence of an event horizon. However, there are three obstacles
one is faced with: (1) the amount of mass that is accreted onto the CMO
is uncertain, because a fraction of the fall-back material could be
blown away from the disk by a radiation driven wind
\citep[e.g.][]{2016MNRAS.461..948M}; (2) as we will show in section
\ref{sec:envelope}, 
for relatively low-mass CMOs ($M< 10^{7.5}M_{\rm\odot}$), the
emission from the stellar debris is mostly in the far UV where either our telescopes are 
currently not sensitive enough or absorption along the line of sight
is strong (for photons with energy $>13.6\rm\ eV$); (3) it is
non-trivial to distinguish the 
emission from the stellar debris from that of the accretion disk.

However, if we consider CMOs more massive than
$10^{7.5} M_{\rm \odot}$, main sequence stars have to get
closer than the innermost stable circular orbit in order to get tidally
disrupted. In such cases, the geodesics in the Schwarzschild
spacetime are plunging (or 
bound), so we expect only a small fraction of the disrupted star to be
blown away and the majority to fall onto the CMO. On the other
hand, when the CMO's mass is so large that its radius is larger than
$r_{\rm T}$ given by eq. (\ref{eq:1}), classical TDEs do not happen and
there is no disk formation (though the star may be disrupted by
relativistic tidal forces if 
it gets very close to $r_{\rm S}$). There are then two possibilities:  (1)
if the CMO has a hard surface, the stellar gas is shocked when
colliding with the surface and the shocked gas forms a hot
radiation-dominated envelope; (2) if the CMO is a diffuse cluster of
particles or very low mass BHs, after entering the CMO, the star (if
not tidally disrupted) experiences a drag due to  dynamical friction
or collisions with the 
particles. Unfortunately, the drag may be too small to 
affect the stellar orbit because the particles could be weakly
interacting with a very small collisional cross section. For instance,
very low mass BHs penetrate through the star at high speed without
producing much friction. There may not be any observational
consequence in the second scenario. Therefore, we only consider the
first scenario, namely, the CMO has a hard surface; some of the
observational constrains we discuss should also apply to the situation
where the star is tidally disrupted inside the CMO if it is dense
enough or consists of 
massive compact objects that are capable of causing disruption.

In this paper, we consider CMOs heavier than
$10^{7.5} M_{\rm \odot}$ with a hypothetical hard
surface at radius $r_0 = \eta r_{\rm S}$ (eq.~\ref{eq:9}), with $1<\eta<30$. We assume CMOs
to be non-rotating (or spin parameter $a/M\lesssim 0.2$),
so the spacetime outside the surface is approximately
spherically symmetric. When the star's
orbit has pericenter distance smaller than max($r_{\rm T},
r_0, 4r_{\rm S}$), it is destroyed due to either tidal disruption or
collision with the surface, which we call {\it stellar disruption
  events} in general. 
Note that a parabolic orbit with pericenter distance smaller
than $4r_{\rm S}$ means specific angular momentum less than $2r_{\rm
  S}c$, so the geodesic in a Schwarzschild metric is  
plunging. In a stellar disruption event, the stellar gas gets shocked
and then forms a quasistatic envelope 
supported by radiation pressure above the surface. In section
\ref{sec:envelope}, we show that the radiation from the stellar debris is
bright at optical/UV wavelengths and could be detected as unique
and long-lasting transients. In section \ref{sec:rate}, we show that,
given the estimated rate of such stellar disruption
events, non-detection of such transients by current optical surveys
has already ruled out a hard surface with $\eta-1\gtrsim10^{-4.4}$. 

\section{Thermal Radiation from the Stellar
  Debris}\label{sec:envelope}  
\subsection{Pressure profile in the strong-gravity
  regime}\label{sec:profile} 

\begin{figure}
  \centering
\includegraphics[width = 0.5 \textwidth,
  height=0.26\textheight]{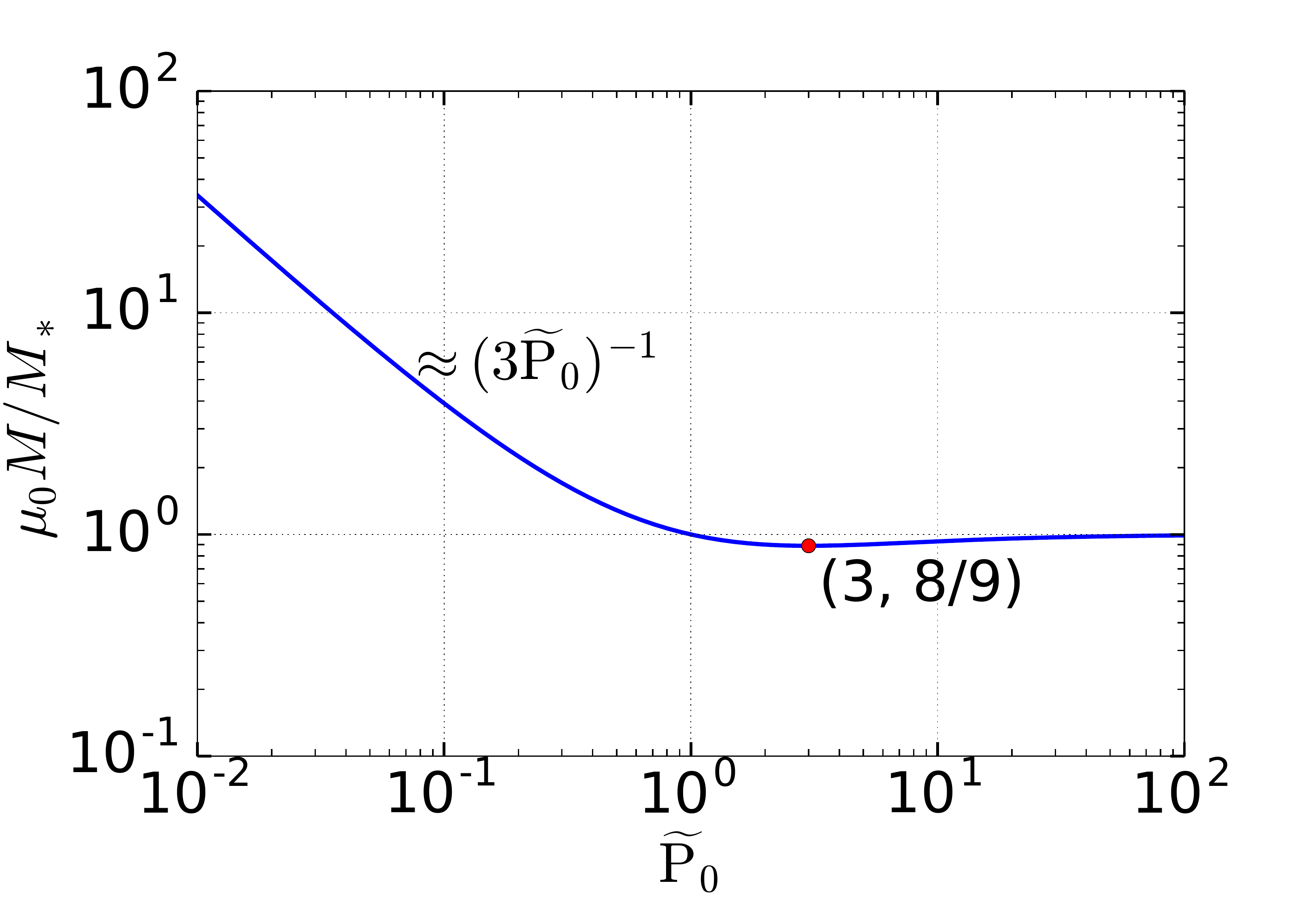}
\caption{The relation between the normalized peak pressure
  $\tp_0$ and the compactness of the horizonless object $\mu_0$, given
  by eq.~(\ref{eq:1022}). When $\mu_0 M/M_* \geq 1$
or $\mu_0 M/M_*=8/9$, the peak pressure is unique for each $\mu_0
M/M_*$; when $8/9<\mu_0 M/M_*<1$, each $\mu_0 M/M_*$
corresponds to two different peak pressures (the solution
corresponding to the larger pressure is unstable); when $\mu_0
M/M_*<8/9$, the TOV equation with a relativistic EoS $P = \rho/3$ has no
solution because, to support gravity, a static configuration
requires a local sound speed greater than $c$.
}\label{fig:p0}
\end{figure}

In this subsection, we use geometrized units $G = c = 1$. Consider a
horizonless object of mass $M$ with a hard  
surface at $r_0$ larger than $r_{\rm S}\equiv2M$. The
compactness of the object is characterized by
\begin{equation}
\label{eq:mu0}
\mu_0 = 1 - \frac{1}{\eta} =
1 - \frac{r_{\rm S}}{r_0}.
\end{equation} 
In this 
subsection we assume $\mu_0\ll 1$, so we have 
\begin{equation}
\label{eq:mu0approx}
\mu_0 \approx \eta - 1 = \frac{r_0}{r_{\rm S}} - 1
\quad{\rm(for}~\mu_0\ll1).
\end{equation} 
When a star of mass
$M_*$ falls onto this object from infinity, gas particles move radially inward
with Lorentz factor $\mu_0^{-1/2}\gg1$ in the local frame before being
shocked at the 
surface. Therefore, the shocked gas is highly relativistic with
equation of state (EoS) $P = \rho /3$ ($P$ is pressure and $\rho$ is
energy density in the fluid rest frame). Note that here both pressure
and energy density are dominated by radiation. We assume the system to
be spherically symmetric. When the system reaches hydrostatic
equilibrium, as long as the matter-radiation mixture can be considered
as a tightly coupled single fluid system with an isotropic pressure
tensor (see Appendix A for more details), the pressure profile of
the shocked gas on the object's surface is described by the
Tolman-Oppenheimer-Volkoff (TOV) equation,
\begin{equation}
  \label{eq:104}
  \frac{\d P}{\d r} = - \frac{(\rho + P)(m + 4\pi r^3 P)}{r(r - 2m)} =
  - \frac{4P(m + 4\pi r^3 P)}{r(r - 2m)},
\end{equation}
where
\begin{equation}
  \label{eq:109}
  m(r) \equiv \int_{r_0}^r 4\pi r^2 \rho(r) \d r + M= 12\pi \int_{r_0}^r r^2
P(r) \d r + M
\end{equation}
is the total mass within radius $r$. For $r/r_{\rm S} - 1
\ll 1$ and $M_*/M\ll 1$, we can make simplifications, $r\approx r_{\rm
  S}$ and $m(r)\approx M$, everywhere except in the $(r - 2m)$ term.
Defining $\tp(r)\equiv 4\pi r_{\rm S}^3 P(x)/M$ and $x(r) \equiv r - 
2m(r)$, we obtain
\begin{equation}
  \label{eq:1014}
  \d x = \d r (1 - 24\pi r^2 P) = \d r (1 - 3 \tp).
\end{equation}
The TOV equation then becomes
\begin{equation}
  \label{eq:1013}
  \frac{\d \tp}{\d x} = -\frac{2\tp (\tp + 1)}{x(1 - 3\tp)}.
\end{equation}
This simplified TOV equation can be integrated, given the boundary
condition 
$\tp(x_0) = \tp_0$,
\begin{equation}
  \label{eq:1018}
  \frac{(\tp + 1)^4}{\tp} = \frac{(\tp_0 + 1)^4}{\tp_0} \left(\frac{x}{x_0}\right)^2,
\end{equation}
where  $x_0 \equiv r_0 - 2M$. When $\tp \gg 1$, we have $\tp\propto
x^{2/3}$. It can be seen from eq. (\ref{eq:1014}) that, as $r$
increases, both $x$ and $\tp$ decrease rapidly and the system is
barely able to avoid the formation of an event horizon, which means
this layer of stellar debris is unstable. On the other hand, 
when $\tp \ll 1$, the pressure profile is a power-law $\tp \propto
x^{-2} \propto (r-2M)^{-2}$. The pressure at the bottom of the stellar debris
$\tp_0$ is given by mass normalization 
\begin{equation}
  \label{eq:1019}
 M_* = 12\pi \int_{r_0}^{r_1} r^2 P(r) \d r,
\end{equation}
where $r_1$ is the outer boundary where $P$ vanishes\footnote{Strictly
  speaking, $P$ does not vanish at the outer boundary because
  there is always a net outward radiation flux. This only
  affects the very surface layer (where $\tp\ll\tp_0$), and the
  pressure profile (eq.~\ref{eq:1018}) and the pressure at the bottom
  of the stellar debris (eq.~\ref{eq:1022}) are not affected.}. Using the 
new notation, we have
\begin{equation}
  \label{eq:1020}
  M_* = \frac{3}{2}\int_{x_0}^{x_1} \frac{\tp \d x}{1 - 3\tp},
\end{equation}
where $x_1 \equiv r_1 - 2m(r_1)$. We know from
equation~(\ref{eq:1018}) that 
\begin{equation}
  \label{eq:1021}
  \frac{\d x}{x_0} = -\frac{\tp_0^{1/2}}{(\tp_0+1)^2} \frac{(\tp
    +1)(1 - 3\tp)}{2\tp^{3/2}} \d \tp,
\end{equation}
so we can integrate the right-hand side of eq.~(\ref{eq:1020}) and obtain
\begin{equation}
  \label{eq:1022}
  \frac{\mu_0 M}{M_*} = \frac{x_0}{2M_*} = \frac{(\tp_0 + 1)^2}{\tp_0
    (\tp_0 +3)}.
\end{equation}
We show in Fig.~(\ref{fig:p0}) the relation between the normalized
peak pressure $\tp_0$ and $\mu_0 M/M_*$. When $\mu_0 M/M_* \geq 1$
or $\mu_0 M/M_*=8/9$, the peak pressure is unique for each $\mu_0
M/M_*$; when $8/9<\mu_0 M/M_*<1$, each $\mu_0 M/M_*$
corresponds to two different peak pressures (the solution
corresponding to the larger peak pressure is unstable); when $\mu_0
M/M_*<8/9$, the TOV equation with a relativistic EoS $P = \rho/3$ has no
solution\footnote{This is analogous the \citet{Buchdahl59} constraint on the radius
  of a relativistic star, but the difference
  is that in the present case there is a hard surface at the bottom of the
  baryonic gas.} because, to support gravity, a static configuration
requires a local sound speed greater than $c$.

\begin{figure}
  \centering
\includegraphics[width = 0.5 \textwidth,
  height=0.39\textheight]{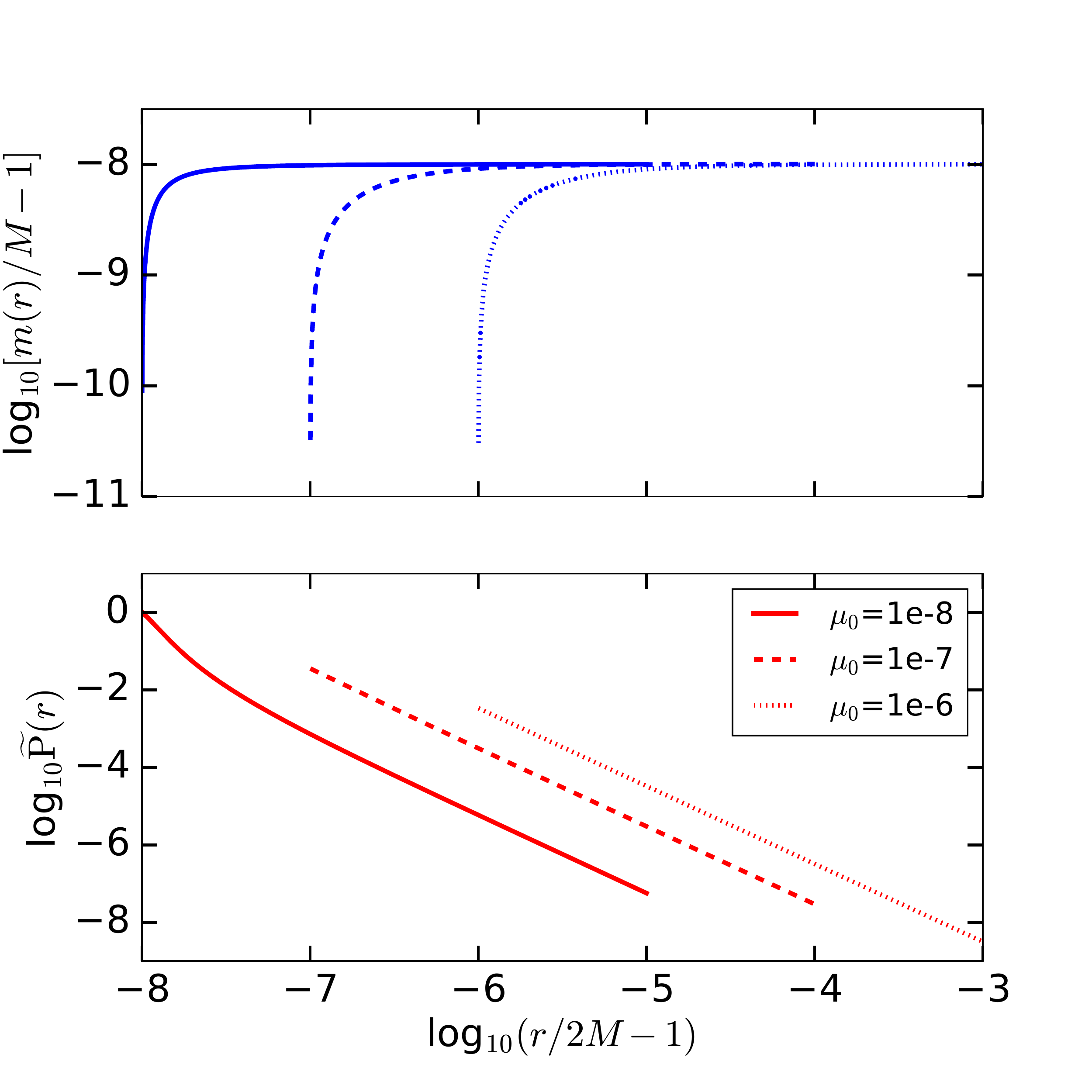}
\caption{The normalized mass and pressure profiles for $M_*/M =
  10^{-8}$ and $\mu_0 = 10^{-8}$ (solid), $10^{-7}$
  (dashed), $10^{-6}$ (dotted).
}\label{fig:profile1}
\end{figure}

Shown in Fig.~(\ref{fig:profile1}) are the mass and pressure profiles
for $M_*/M = 10^{-8}$ and $\mu_0 = 10^{-8}, 10^{-7},
10^{-6}$. The energy density at the
bottom of the stellar debris is (in CGS units)
\begin{equation}
  \label{eq:12}
  \rho(r = r_0) \approx \frac{M_*c^2}{4\pi r_{\rm S}^3\mu_0},
\end{equation}
corresponding to a radiation temperature of $T(r = r_0) \approx
2.9\times10^8 (M_*/M_{\rm \odot})^{1/4} M_8^{-3/4}
\mu_{0,-7}^{-1/4}\rm\ K$ (or 25 keV). At this temperature,
a small fraction of the radiation could be converted to electron-positron
pairs at the bottom of the stellar debris.

Note that it is physically impossible for a horizonless
object with compactness $\mu_0$ to support a layer of stellar debris
of mass 
$M_*>9\mu_0M/8$ with a relativistic EoS $P = \rho/3$. For example, in
the gravastar model, $r_0 - r_{\rm S}$ is on the order of Planck length
$\sim 10^{-33}\rm\ cm$, so the stellar debris has
to switch to the EoS of exotic matter quickly enough to avoid the
formation of an event horizon. To avoid going into details of the
state transition from baryonic to exotic matter, we consider
in this paper only models with $\mu_0\gg M_*/M$.

\subsection{Emission from the photosphere}\label{sec:emission}

In this subsection, we go back to CGS units and discuss the
emission from the stellar debris on the hard surface of the CMO as
viewed by an observer at infinity. We consider the
situation where $\mu_0\gg M_*/M$ (but $\mu_0$ is not necessarily much 
less than 1). The baryonic gas and radiation do not affect the
spacetime outside the hard surface, which is given by the
Schwarzschild metric for a slowly or non-rotating CMO. We define the
function $\mu(r)$ as 
\begin{equation}
  \label{eq:8}
  \mu(r) \equiv -g_{\rm tt}(r) = 1 - \frac{r_{\rm S}}{r},
\end{equation}
where $r_{\rm S}\equiv 2GM/c^2$. When $M_*\sim 1 M_{\rm \odot}$ of
baryonic gas falls onto the CMO in nearly the radial direction, the
gas collides with the 
surface at a locally measured Lorentz factor $\mu_0^{-1/2}$. The
high-density gas downstream of the shock is dominated by radiation
pressure, which is given by
\begin{equation}
  \label{eq:28}
  P_{\rm sh} = \frac{\mu_0^{-1/2} - 1}{3} \rho_{\rm 0, sh}c^2,
\end{equation}
where $\rho_{\rm 0, sh}$ is the local baryonic mass density right after
the shock. Then the stellar debris settles down adiabatically into
quasi-hydrostatic equilibrium. When the spacetime outside the CMO hard
surface is not affected by the existence of the baryonic gas, the
$4\pi r^3 P$ term in the TOV equation can be ignored. The pressure is
dominated by radiation, i.e. $P = P_{\rm rad} + P_{\rm g} \approx
P_{\rm rad}$. We denote the local baryonic mass density as $\rho_0$,
so the total energy density is $\rho = \rho_0 c^2 + 3P_{\rm g}/2 +
3P_{\rm rad}\approx \rho_0 c^2 + 3P$. The TOV equation
can then be simplified to
\begin{equation}
  \label{eq:7}
  \frac{\d P}{\d r} = -\mu(r)^{-1} \frac{GM \rho_0}{r^2} \left(1 +
    \frac{4P}{\rho_0 c^2} \right).
\end{equation}

If the internal energy per baryon is roughly conserved, we
have $P/\rho_0 \simeq P_{\rm sh}/\rho_{\rm 0,sh}$ and the pressure scale
height of the stellar debris at $r = r_0$ is
\begin{equation}
  \label{eq:16}
  H = \mu_0\frac{2r_0^2}{r_{\rm S}}\frac{P_{\rm sh}/(\rho_{\rm 0,sh}
    c^2)}{1 + 4P_{\rm sh}/(\rho_{\rm 0,sh} c^2)} \simeq
  \mu_0\frac{2r_0^2}{r_{\rm S}} \frac{1 - \sqrt{\mu_0}}{4 -
    \sqrt{\mu_0}}.
\end{equation}
Note that $H$ denotes the scale height in Schwarzschild
coordinates; the physical (locally measured) scale height is
$H\mu_0^{-1/2}$. In the limit $r_0\rightarrow r_{\rm S}$ ($\mu_0\ll
1$), we get $H\simeq (r_0-r_{\rm S})/2$, and in the limit $r_0\gg r_{\rm S}$
($\mu_0\approx 1$), we get $H\simeq r_0/3$. In reality, part of the internal
energy is used to do work against ``gravity'' (when
$\mu_0\gtrsim10^{-6}$, we have $H\mu_0^{-1/2} \gtrsim$ solar radius), so
the scale height will be 
smaller (but this has little effect on the analysis since we already have
$H<r_0-r_{\rm S}$). In addition, part of the internal energy could be
used to drive a wind, which might carry a fraction of the total mass
$M_*$ away at the local escape velocity, so the scale height will
be even smaller. We note that the fractional wind mass loss is small
($\ll1$) because the extra energy taken away by the wind makes the
rest of the gas even more bound. We conclude that the pressure or
density scale height is roughly a factor of a few smaller than 
$r_0-r_{\rm S}$.

If the total mass of the stellar debris is $M_{*} = \xi M_{\rm \odot}$
($\xi\lesssim1$), the Thomson depth of the whole layer is
\begin{equation}
  \label{eq:29}
  \tau_0 \simeq \frac{\kappa_{\rm T}M_{*}}{4\pi r_0^2} \simeq
  6.0\times 10^4 \xi \eta^{-2} M_8^{-2},
\end{equation}
where we have used the Thomson opacity for solar metallicity,
$\kappa_{\rm T} = 0.34\rm\ cm^2\ g^{-1}$. The photospheric radius
$r_{\rm ph}$ (or $\mu_{\rm ph}\equiv \mu(r_{\rm ph})$) where the Thomson 
scattering optical depth is order unity is larger than $r_0+H$ (due to
the large total optical depth). For an observer at infinity, the
diffusion time across the entire layer of stellar debris is roughly
given by
\begin{equation}
  \label{eq:30}
  \begin{split}
      t_{\rm dif,\infty} &\simeq \tau_0 \int_{r_0}^{r_{\rm
      ph}}\frac{\d r}{c}\mu^{-1} \simeq \frac{\tau_0 \eta r_{\rm
      S}}{c}.
  \end{split}
\end{equation}
During a time $t_{\rm dif,\infty}$, an amount of radiation energy
(viewed at infinity) $(1- \mu_0^{1/2})M_*c^2$ diffuses outwards, which gives
a diffusive luminosity
\begin{equation}
  \label{eq:42}
  L_{\rm dif, \infty} \simeq \frac{(1- \mu_0^{1/2})M_*c^2}{t_{\rm
      dif,\infty}} \simeq (1- \mu_0^{1/2}) \eta^{-1} L_{\rm Edd},
\end{equation}
where the Eddington luminosity is $L_{\rm
  Edd}\equiv 4\pi c G M/\kappa_{\rm T} = 1.5\times10^{46}M_8\rm\ erg\
s^{-1}$. From eq.~(\ref{eq:42}), we find that $L_{\rm dif,\infty}\simeq
L_{\rm Edd}$ when either $\mu_0\ll1$ or
$\mu_0\approx1$. Including gravitational redshift, the diffusive
flux in the local rest frame at $\mu(r)$ is $\simeq L_{\rm
  Edd}\mu^{-1}$, which means that the radiation 
force on the baryon-photon mixture balances gravity\footnote{The
  gravitational acceleration in the local rest frame at $r$ is
  $GM\mu^{-1/2}/r^2$ and each electron has an effective inertia
  $\mu^{-1/2} m_{\rm p}$ (dominated by radiation).}.

Photons emitted at the photosphere at radius $r_{\rm ph}$, or
$\mu_{\rm ph}\equiv \mu(r_{\rm ph})$, may not escape to
infinity. The maximum polar angle
$\theta_{\rm m}$ up to which photons emitted at $r_{\rm ph}$ can escape to infinity is given by
\begin{equation}
  \label{eq:40}
  \begin{split}
       \theta_{\rm m} =  & \pi/2\mbox{, if $r_{\rm ph} > 1.5 r_{\rm S}$,}\\
    \theta_{\rm m} = & \mathrm{sin}^{-1}
      \frac{3\sqrt{3}\mu_{\rm ph}^{1/2}}{2r_{\rm ph}/r_{\rm
      S}} \mbox{, if $r_{\rm ph} \leq 1.5 r_{\rm S}$.}
  \end{split}
\end{equation}
The luminosity seen by an observer at infinity is the fraction of
$L_{\rm dif,\infty}$ that escapes, i.e.
\begin{equation}
  \label{eq:340}
      L_{\rm \infty}  = L_{\rm dif,\infty}\frac{\int_0^{\rm \theta_{\rm
            m}} I(\theta) \cos\theta \sin \theta\d \theta}{
        \int_0^{\pi/2} I(\theta) \cos\theta \sin \theta\d \theta},
\end{equation}
where $L_{\rm dif,\infty} = 8\pi^2r_{\rm ph}^2\int_0^{\pi/2} I(\theta) \cos\theta
\sin \theta\d \theta \simeq L_{\rm Edd}$. The intensity is
nearly angle independent $I(\theta)\simeq I$, so we have
\begin{equation}
  \label{eq:34}
  L_{\rm \infty}\simeq L_{\rm Edd}\sin^2\theta_{\rm m}.
\end{equation}
For any given angle, the spectrum is nearly a blackbody $I_{\nu}
\simeq B_{\nu}(T_{\rm ph})$ \citep[e.g.][]{2006ApJ...638L..21B}, so we
have $I\simeq \sigma_{\rm SB} T_{\rm ph}^4/\pi$, where $\sigma_{\rm
  SB}$ is the Stefan-Boltzmann constant and $T_{\rm ph}$ 
is the local radiation temperature at the photosphere. Therefore, the
radiation temperature at infinity is
\begin{equation}
  \label{eq:33}
  \begin{split}
              T_{\rm \infty} &= T_{\rm ph} \mu_{\rm ph}^{1/2} =
          \left(\frac{L_{\rm Edd}}{4\pi r_{\rm ph}^2\sigma_{\rm
                SB}} \right)^{1/4}\mu_{\rm ph}^{1/4}\\
          &\simeq 3.9\times10^5 \left(\frac{r_{\rm S}}{r_{\rm
                ph}}\right)^{1/2} M_8^{-1/4} \mu_{\rm ph}^{1/4}\rm\ K.
  \end{split}
\end{equation}
The duration of the emission from the stellar debris is given by
energy conservation
\begin{equation}
  \label{eq:43}
      \Delta t_{\rm \infty} = (1- \mu_0^{1/2})\frac{\xi M_{\rm
      \odot}c^2}{L_{\rm \infty}} \simeq 1.2\times 10^8 \frac{\xi (1-
    \mu_0^{1/2})}{M_8 
    \sin^2\theta_{\rm m} }\rm\ s. 
\end{equation}
Note that in the limit $\mu_{\rm ph}\ll 1$ (and $\eta\approx1$), the
duration of the transient emission in eq.~(\ref{eq:43}) 
is longer than the diffusion time given by eq.~(\ref{eq:30}) by a
factor $\mu_{\rm ph}^{-1}$. This is because photons emitted at the
photosphere tend to be lensed back $\mu_{\rm ph}^{-1}$ times before
escaping (a photon can escape only when $\theta\leq\theta_{\rm m}$).

\begin{figure}
  \centering
\includegraphics[width = 0.5 \textwidth,
  height=0.21\textheight]{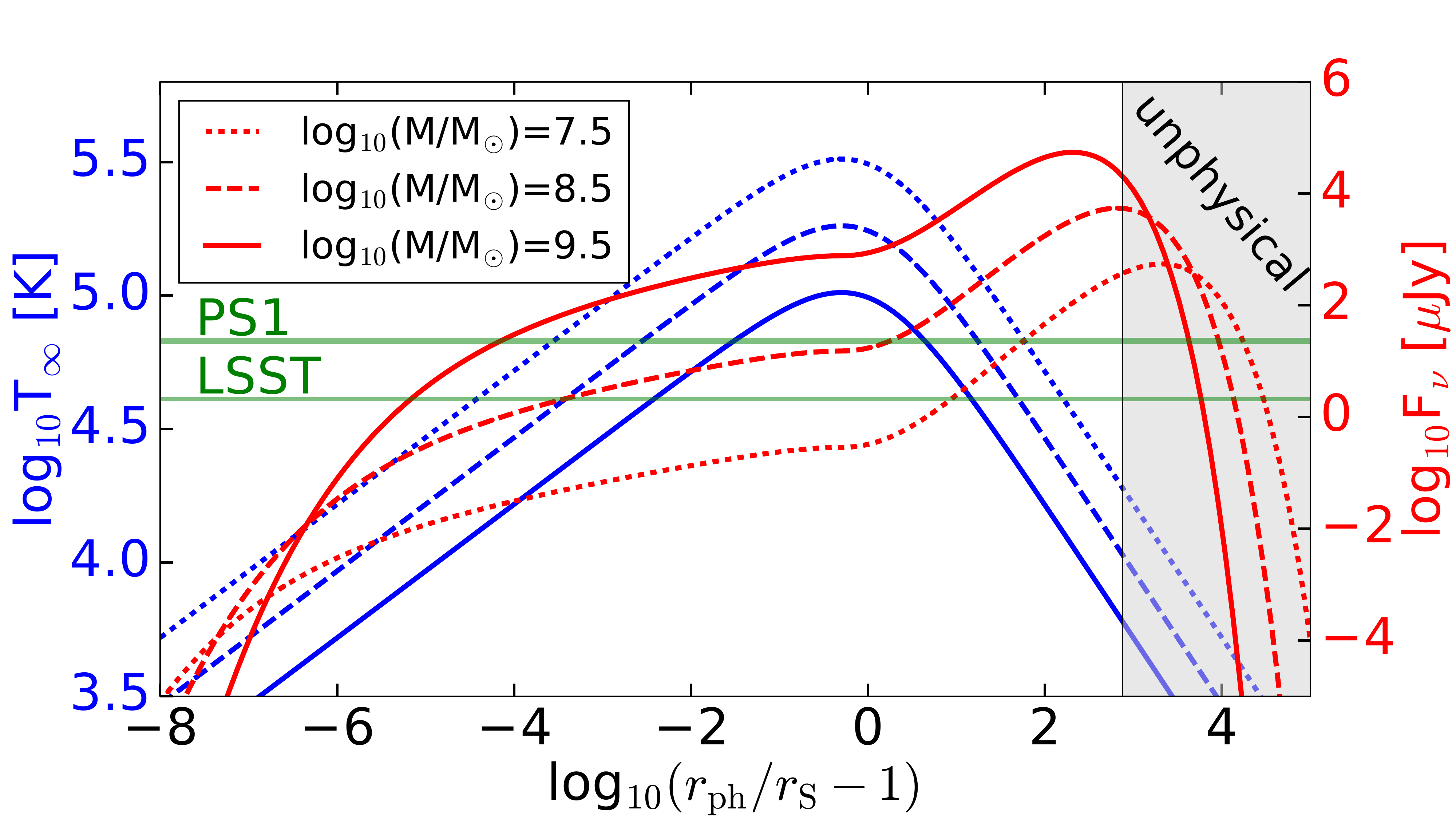}
\includegraphics[width = 0.5 \textwidth,
  height=0.25\textheight]{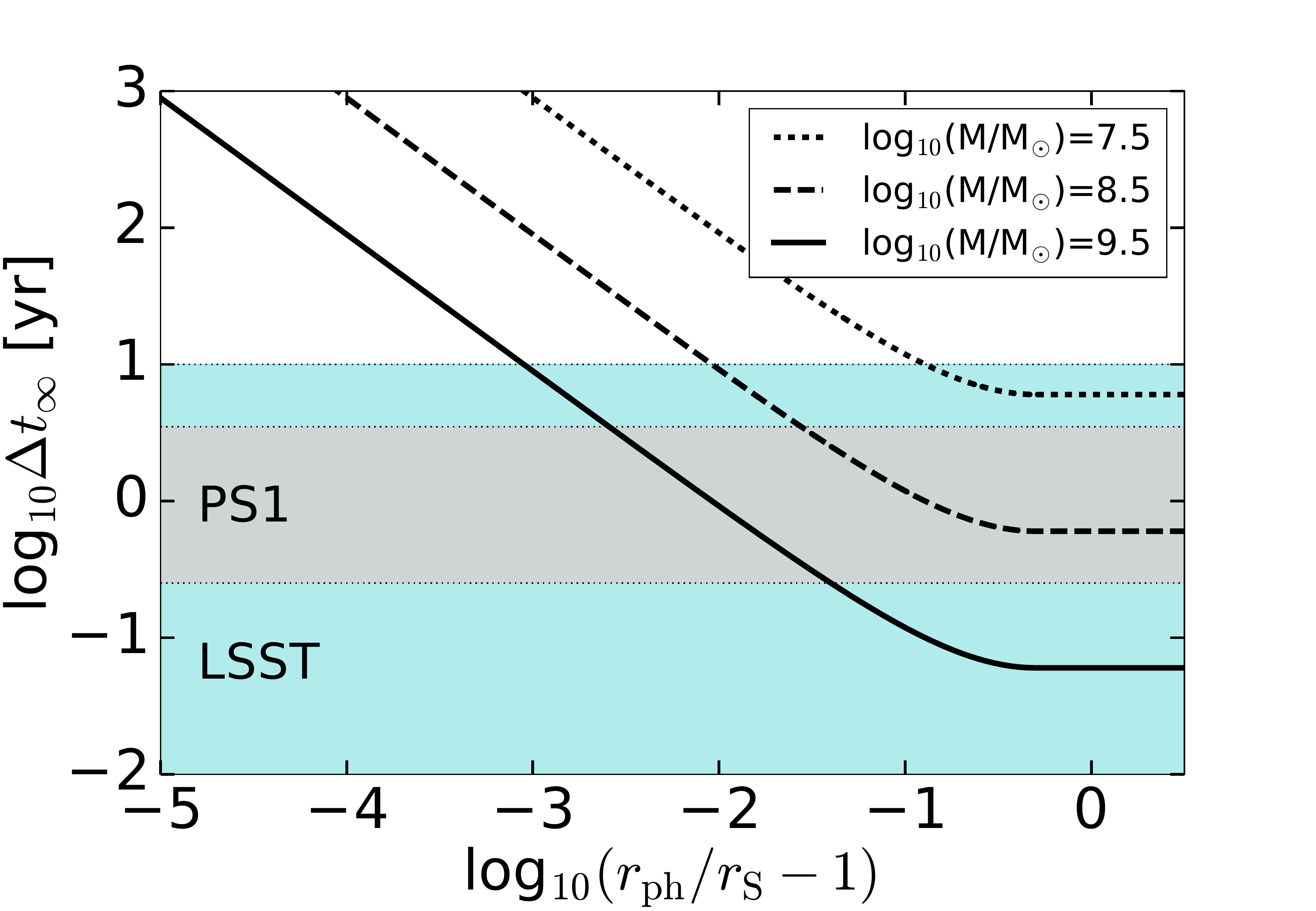}
\caption{For an observer at infinity, the emission from the layer of
  stellar debris on the hard surface has a blackbody spectrum with
  luminosity given by eq.~(\ref{eq:34}) and temperature given by
  eq.~(\ref{eq:33}). The duration of the emission is given by
  eq.~(\ref{eq:43}). In the {\it Upper Panel}, we show in red lines
  the g-band flux density as a function of the photospheric radius
  $r_{\rm ph}$, for three different CMO masses, $M = 10^{7.5}\rm\
  M_{\rm\odot}$ (dotted), $10^{8.5} \rm\ M_{\rm\odot}$ (dashed)
  and $10^{9.5}\rm\ M_{\rm\odot}$ (solid), at redshift $z=0.5$. The
  limiting flux for Pan-STARRS1 3$\pi$ survey (PS1, thick green line)
  and LSST (thin green line) are calculated by assuming the source to
  be 1.5 mag brighter than the 5$\sigma$ sensitivity for a single 
  exposure. For CMOs heavier than
  $10^{7.5}M_{\rm \odot}$, the photospheric radius must be smaller
  than $7.6\times10^2r_{\rm S}$ (otherwise the entire baryonic layer
  is Thomson thin), so the grey shaded region is unphysical.
  We also show in blue lines the radiation temperature for an
  observer at infinity as a function of $r_{\rm ph}$. In the {\it
    Lower Panel}, we show 
  the duration of the emission $\Delta t_{\infty}$ as a function of
  $r_{\rm ph}$ when the total mass of the stellar debris is $M_* =
  0.5M_{\rm \odot}$ ($\xi = 0.5$). The grey/blue
  regions denote the timescales over which PS1/LSST are complete, with
  the upper bound given by the survey lifespan and the lower bound given
  by the cadence (see section \ref{sec:rate} for details of the two
  telescopes). We only show the parameter space $r_{\rm 
    ph}/r_{\rm S} - 1 \lesssim 1$, because in this regime $\Delta
  t_{\infty}$ is only a function of $r_{\rm ph}$ but not $r_0$ (since 
  $\mu_0\approx1$). As we show in section \ref{sec:rate}, the
  parameter space $r_{\rm ph}/r_{\rm S} - 1 \gtrsim1$ has already been
  ruled out by PS1.
}\label{fig:emission}
\end{figure}

In the upper panel of Fig.~(\ref{fig:emission}),
we show in red lines the g-band ($\simeq 4800\rm\ \AA$) flux
density as a function of the photospheric radius $r_{\rm ph}$, for
three different  
CMO masses, $M = 10^{7.5}$, $10^{8.5}$ and $10^{9.5}\rm\ M_{\rm\odot}$, at
redshift $z=0.5$. The limiting flux for the Pan-STARRS1 3$\pi$ survey
(PS1, thick green line) and the Large Synoptic Survey Telescope 3$\pi$
survey (LSST, thin 
green line) are calculated by assuming the source to be at least 1.5
mag brighter than the 5$\sigma$ 
flux limit for a single exposure ($M_{\rm AB}=22.0$ and 23.4 mag for
PS1 and LSST 
respectively, see section \ref{sec:rate} for details of the two
telescopes). We also show in red lines the blackbody
temperature for an observer at infinity as a function of $r_{\rm ph}$. 

When $r_{\rm ph}/r_{\rm S} - 1\ll1$, we have $\mu_{\rm ph} \approx
r_{\rm ph}/r_{\rm S} - 1$, $L_{\rm \infty}\propto \mu_{\rm
  ph}$ and $T_{\infty}\propto \mu_{\rm ph}^{1/4}$. For
$10^{-5}\lesssim r_{\rm ph}/r_{\rm S} - 1\ll1$, the g-band frequency
is in the Rayleigh-Jeans regime, so the flux density decreases
slowly as $F_{\nu}\propto \mu_{\rm ph}^{1/4}$. For
$\mu_{\rm ph}\lesssim 10^{-5}$, the g-band frequency slowly shifts 
into the Wien tail, so the flux density drops faster. On
the other hand, for $r_{\rm ph}/r_{\rm S} - 1\gg1$, the temperature
decreases rapidly as $T_{\infty}\propto r_{\rm ph}^{-1/2}$, but the
luminosity stays constant at $L_{\rm Edd}$. Therefore, the flux
density first increases 
as $F_{\rm \nu} \propto r_{\rm ph}^{3/2}$ in the Rayleigh-Jeans regime and
then decreases exponentially in the Wien regime.

Note that the optical depth for Thomson scattering at the photosphere  
is of order unity, so this gives an upper limit for the
photospheric radius,
\begin{equation}
  \label{eq:35}
  4\pi r_{\rm ph}^2 < \kappa_{\rm T}M_{*},
\end{equation}
which means
\begin{equation}
  \label{eq:36}
  \frac{r_{\rm ph}}{r_{\rm S}}<7.6\times10^2\xi^{1/2} M_{7.5}^{-1}.
\end{equation}
For $\xi \lesssim 1$ and the CMO mass range
$M>10^{7.5}M_{\rm \odot}$  considered in this paper, the photospheric radius of the stellar debris 
layer must be smaller than $7.6\times10^2r_{\rm S}$, so the grey
shaded region in the upper panel of Fig.~(\ref{fig:emission}) is
unphysical. For CMOs with mass $M<10^{7.5}M_{\rm
  \odot}$, the emission from the stellar debris may peak in the
non-observable far UV (if $r_{\rm 
  ph}/r_{\rm S} - 1\lesssim1$), so one does not obtain strong constraints
on the radius of the hard surface. In addition, stars get tidally disrupted
before reaching close to the Schwarzschild radius, so the 
actual accretion rate is uncertain due to the complexities of
accretion disk physics.

In the lower panel of Fig.~(\ref{fig:emission}), we show the emission
duration $\Delta 
t_{\infty}$ as a function of $r_{\rm ph}$ for the same stellar mass
$M_* = 0.5M_{\rm \odot}$ ($\xi = 0.5$) and three CMO
masses. From eq.~(\ref{eq:43}), we see that $\Delta t_{\infty}$ is a
function of both the hard surface radius $r_0$ (or $\mu_0$) and
the photospheric radius $r_{\rm ph}$. As we show in section
\ref{sec:rate}, observations from PS1 have already ruled out the
parameter space $r_{\rm ph}/r_{\rm S} - 1 \gtrsim 1$, so here we only
show the parameter space $r_{\rm ph}/r_{\rm S} - 1 
\lesssim 1$, where $\Delta t_{\infty}$ is only a simple function of
$r_{\rm ph}$ (because $\mu_0\ll 1$).
We note that $\Delta t_{\rm \infty}\propto \mu_{\rm ph}^{-1}$ in
the limit of $\mu_{\rm ph} \ll 1$. Therefore, if the CMO
is compact enough, the duration may become longer than the average
time interval between two stellar disruption events and persistent
emission from CMOs could be searched for\footnote{This is similar to
  what has been done on Sgr A*, M87* and BH candidates in some X-ray  
  binaries (see \citealt{nar97, 2007CQGra..24..659B, 2008NewAR..51..733N, 2009ApJ...701.1357B,
    2015ApJ...805..179B}). The persistent emission is most likely
  dominated by the gas accreted in the AGN phase instead of stellar
  disruption events, because the former dominates CMOs' mass growth.}.

To link the observational constraints on the photospheric radius
$r_{\rm ph}$  to the physical limits on the hard surface radius
$r_0$, we need to calculate the baryonic density profile of the
stellar debris layer. The detailed density profile of the 
stellar debris could in principle be obtained by considering the radiation
transfer and gravity with appropriate EoS and boundary conditions
\citep[e.g.][]{1986ApJ...302....1P, 2016MNRAS.458.3420W}. For the
purpose of this paper, we only need to consider the baryonic 
density profile in the optically thick region in the limit
$\mu_{\rm ph}\ll1$. The system reaches hydrostatic equilibrium roughly
on the light-crossing timescale $\sim r_{\rm S}/c$ (or a logarithmic
factor larger), which is much smaller than the diffusion time, so the
evolution of the stellar debris can be considered as adiabatic and we
have $P(\mu)\propto [\rho_0(\mu)]^{4/3}$ (it is more convenient  
to use $\mu$ instead of the radial coordinate $r$). From section
\ref{sec:profile}, we know $P(\mu)\propto \mu^{-2}$, so the baryonic
density profile is $\rho_0(\mu) \propto \mu^{-3/2}$. The normalization
is given by the total mass $M_* = \int_{r_0}^{r_1}
4\pi r^2 \rho_0(r) \mu^{-1/2}\d r$, so we have 
\begin{equation}
  \label{eq:10}
  \rho_0(\mu) = \frac{M_*}{4\pi r_{\rm S}^3 \mu_0^{1/2}}
  \left( \frac{\mu}{\mu_0} \right)^{-3/2}. 
\end{equation}
From the rest mass density and temperature (eq. \ref{eq:12}), we get
the ratio between the (non-relativistic) electron degeneracy pressure
and gas pressure $P_{\rm deg}/P_{\rm g}\simeq 2\times10^{-7} 
  (M_*/M_{\rm \odot})^{5/12} M_8^{-5/4} \mu_{0,-7}^{-1/12}
  (\mu/\mu_0)^{-1/2}$. The ratio between gas pressure and
  radiation pressure is $P_{\rm g}/P_{\rm rad}\simeq 10^{-8} 
 (M_*/M_{\rm \odot})^{1/4} M_8^{-3/4} \mu_{0,-7}^{1/4}$. Therefore,
 the pressure is completely dominated by radiation.
The Thomson optical depth above a certain radius $\mu(r)$ is
\begin{equation}
  \label{eq:11}
      \tau(\mu) = \frac{\kappa_{\rm T}}{4\pi r_{\rm S}^2} \int_r^{r_1} 4\pi
  r^2 \rho_0(r) \mu^{-1/2} \d r =\frac{\kappa_{\rm T}M_*}{4\pi r_{\rm S}^2}
  \frac{\mu_0}{\mu}.
\end{equation}
Therefore, the relation between the photospheric radius $r_{\rm ph}$
and the hard surface radius $r_0$, in the
limit $\mu_{\rm  ph}\ll 1$, is 
\begin{equation}
\label{eq:muph}
\mu_{\rm ph}\simeq \mu_0
\frac{\kappa_{\rm T}M_*}{4\pi r_{\rm S}^2}.
\end{equation}

For a given CMO mass $M$, as long as $\mu_0> 4\pi r_{\rm
  S}^2/(\kappa_{\rm T}M_*)$, the photospheric radius is at $r_{\rm
  ph}/r_{\rm S} - 1\gtrsim 1$. As 
shown in Fig.~(\ref{fig:emission}), the g-band flux density increases
roughly as $r_{\rm ph}^{3/2}$ in this regime. Instead of solving for the detailed
baryonic density profile when $r_{\rm ph}/r_{\rm S} - 1\gtrsim 1$, we
take a conservative limit\footnote{We are taking $r_{\rm ph} =
  \mbox{max}(1.3r_{\rm S}, r_0)$ when $r_{\rm ph}/r_{\rm S} - 1\gtrsim
  1$. Since the flux density increases with $r_{\rm ph}$ (see
  the upper panel of Fig.~\ref{fig:emission}) while the
  duration of the transient emission is nearly not affected by $r_{\rm ph}$
  (eq.~\ref{eq:43}), the actual detectable event rate for a given
  survey is higher than suggested by our calculations.}
\begin{equation}
  \label{eq:14}
  \frac{r_{\rm ph}}{r_{\rm S}} - 1 = \mbox{min}\left[\mu_0\tau_0,\
  \mbox{max}\left(0.3,~\frac{r_0}{r_{\rm S}}-1\right)\right], 
\end{equation}
where $\tau_0 = \kappa_{\rm T} M_*/(4\pi r_0^2)\simeq
6.0\times10^4 \xi\eta^{-2} M_8^{-2}$.
Eq.~(\ref{eq:14}) and $\mu_{\rm ph} = 1 - r_{\rm S}/r_{\rm
  ph}$, as well as eqs.~(\ref{eq:34}), (\ref{eq:33}) and (\ref{eq:43}), will be
used in section \ref{sec:rate} to calculate a  
lower limit on the observed the flux density for a CMO of given mass at a given
redshift. One more point to note is that, when considering
$\eta\gg1$, we discard the (very few) high mass CMOs that give $\tau_0 
\leq 10$, to make sure that the radiation field is well
thermalized. 
\section{Observations}\label{sec:rate}

In this section, we assume the total baryonic mass of the stellar
debris layer to be $M_* = \xi M_{\rm \odot} = 0.5 M_{\rm \odot}$. For
a given CMO of mass $M$ and redshift $z$, the flux density at
frequency $\nu$ on 
the Earth is
\begin{equation}
  \label{eq:17}
  F_\nu = \frac{15}{\pi^4} \frac{L_{\rm \infty}}{4\pi D_{\rm L}^2}
  \frac{x^4/\nu}{\mathrm{e}^x-1},
\end{equation}
where $x = h\nu(1+z)/(kT_{\rm \infty})$, $h$ is the Planck constant,
$k$ is the Boltzmann constant, and $D_{\rm L}(z)$ is the luminosity
distance\footnote{We use a
  standard $\Lambda$ cold dark matter cosmology with $H_{\rm 0} = 70
  \rm\ km\ s^{-1}\ Mpc^{-1}$, $\Omega_{\rm m} = 0.27$, and
  $\Omega_{\Lambda} = 0.73$.}. For a survey with limiting flux $F_\nu^{\rm 
  lim}$, we can calculate the limiting redshift $z^{\rm lim}$ by solving
$F_\nu(z) = F_\nu^{\rm lim}$. If we know the mass function of CMOs,
$\Psi(M, z)$ (comoving number density of CMOs of 
different masses at a given redshift), we can calculate the expected
detectable event rate within a solid angle 
$\Delta \Omega$ on the sky
\begin{equation}
  \label{eq:23}
  \dot{N}_{\rm det} = \int_{M_{\rm min}}^{M_{\rm max}}\mt{d} M \dot{N}(M) 
  \int_0^{z_{\rm lim}}\mt{d}z \Psi(M, z) \frac{\mt{d}V}{\mt{d}\Omega
    \mt{d} z} \Delta \Omega,
\end{equation}
where $\dot{N}(M)$ is the stellar disruption rate for a given CMO
of a certain mass, $M_{\rm min} = 10^{7.5}M_{\rm\odot}$ is the minimum
mass we consider, $M_{\rm max}=10^9M_{\rm \odot}$ is the maximum
mass\footnote{We choose $M_{\rm max}=10^9M_{\rm \odot}$ because CMO
  mass function models have too large uncertainties above this
  mass. Since CMO mass functions drop rapidly above
  $10^9M_{\rm \odot}$, our results are not sensitive to
  $M_{\rm max}$. We also tried $M_{\rm max}=10^{9.5}M_{\rm \odot}$ and
the differences are negligible.} we consider, and $\mt{d}V/(\mt{d}\Omega \mt{d} z)$ is the
comoving volume per unit redshift per steradian. 

We use the CMO mass function $\Psi(M,z)$ by
\citet{2009ApJ...690...20S}, who integrate from the low-redshift CMO mass
function backwards over cosmic time with the growth/accretion rate
empirically derived from AGN luminosity function and a prescribed
radiation efficiency. We ignore the (small) contribution from CMOs at
$z>5$, due to large uncertainties on the mass function at high
redshift. We have also tried the mass 
function given by \citet{2008MNRAS.388.1011M}, 
who use the same method as \citet{2009ApJ...690...20S}, and the
differences are negligible. The CMO mass function can also be derived by 
linking their growth to the properties of host dark matter haloes. For
instance, in \citet{2008ApJS..175..356H}, the CMO masses are assumed to be
proportional to the host spheroidal mass, as the host dark matter
haloes grow through major mergers. Various CMO mass function models
are reviewed by \citet{2012AdAst2012E...7K}. At redshift $z<5$, they
agree to within a factor $\lesssim 3$ in the range $10^{7.5}$-$10^{9}
M_{\rm\odot}$ and they all have rapid drop-offs above $\sim 10^{9}
M_{\rm\odot}$.

If we know $\Psi(M, z)$ well enough, the question comes down to the
stellar disruption rate $\dot{N}$ per CMO, which is defined as the sum
of the 
rates of the following three possibilities: (1) the star passes within
the tidal disruption radius $r_{\rm T}$; (2) the specific angular
momentum of the orbit is less than $2r_{\rm S}c$ (corresponding to
a Newtonian parabolic orbit with pericenter distance of $r_{\rm p} =
4r_{\rm S}$); 
(3) the star directly collides with the surface at radius $r_0$.
These rates depend on the stellar phase-space distribution and the
galactic gravitational potential (and other factors mentioned in section
\ref{sec:intro}). If various CMO-host-galaxy correlations
\citep[e.g.][]{2013ARA&A..51..511K} are used, such as 
$M$-$\sigma$ (velocity dispersion) and $M$-$L_{\rm bulge}$ (bulge
luminosity), we can quantify the stellar disruption rate $\dot{N}$
purely as a function of the CMO mass.

The disruption rate has
been extensively calculated for different samples of elliptical
galaxies \citep[e.g.][]{1999MNRAS.309..447M,
  2004ApJ...600..149W, 2016MNRAS.455..859S}. We note that previous
authors chose the critical pericenter distance to 
be $r_{\rm p}= r_{\rm T}$, so when $r_{\rm  T}<\mathrm{max}(r_0, 4r_{\rm
  S})$, the size of the ``loss cone'' and hence the disruption rate were 
underestimated. However, for a given CMO mass and 
stellar phase-space distribution, $\dot{N}$ depends weakly
on the critical pericenter distance $r_{\rm p}$ (roughly as $r_{\rm
  p}^{1/4}$), so the error on the derived disruption rate is small.

Typically, the disruption rate per CMO as a function of the
CMO mass can be described as a power-law,
\begin{equation}
  \label{eq:22}
  \dot{N} = \dot{N}_0 M_{6.5}^{-\delta},
\end{equation}
but the parameters $\dot{N}_0$ and $\delta$ depend strongly on
the galaxy sample. There is a bimodal distribution of
central surface brightness profiles in early-type galaxies
\citep[e.g.][]{2007ApJ...664..226L}. The disruption rates in cusp galaxies
(brightness power-law index $\gamma > 0.2$) are a factor of $\sim 
10$ higher than in core galaxies ($\gamma < 0.2$) with the same CMO
mass. The power-law indexes $\delta$ derived from only cusp or core
galaxies in \citet{2007ApJ...664..226L} are $\delta\simeq 0.25$, but the
power-law is significantly steeper, $\delta\sim 0.4$-0.5, when
the entire sample is considered \citep{2016MNRAS.455..859S}. This is
because core galaxies (with lower $\dot{N}$) generally host more
massive CMOs than cusp galaxies (with larger $\dot{N}$). Other factors,
e.g. non-spherical and time-dependent galactic potential, binary CMOs,
massive perturbers, add further uncertainties on the disruption
rates. It is currently not possible to calculate the
disruption rates as a function of CMO mass \citep[for 
recent discussions, see][]{2013ApJ...774...87V, 2013CQGra..30x4005M,
  2016MNRAS.461..371K}.

On the observational side, several dozen TDE flares have recently been
discovered in surveys from optical to X-ray wavelengths, and the
TDE rate is found to be $\sim10^{-5} \rm\ galaxy^{-1} yr^{-1}$
\citep{2002AJ....124.1308D, 2008ApJ...676..944G, 
    2012ApJ...749..115W, 
    2014ApJ...792...53V, 2016MNRAS.455.2918H}. As pointed out by
\citet{2016MNRAS.455..859S}, there is a 
factor of $\sim 10$ disagreement between the observational and
theoretical TDE rates, which could be due to either observational
incompleteness (e.g. dust extinction or incomplete wavelength
coverage), over-estimate of the brightness of most TDEs
\citep[e.g.][]{2015ApJ...809..166G}, or missing physics in TDE rate
calculations (e.g. time-dependent gravitational potential).
\citet{2016ApJ...818L..21F} show that optical-UV TDEs favor post 
starburst galaxies with CMO mass in the range $10^{5.5}$-$10^{7.5}\rm\
M_{\rm \odot}$, and hence normal star-forming and early-type galaxies
may have a much lower TDE rate. This makes the tension between 
observational and theoretical TDE rates even stronger. Larger samples
in the future will help to illuminate this puzzle.

In the following, we take a conservative estimate for the
  observed TDE rate, $\dot{N}_0 =
1\times10^{-5} \rm\ M_{\rm\odot}\ yr^{-1}$, and leave
the power-law index $\delta\in [0.2, 0.5]$ as a free parameter.
With the detectable event rate $\dot{N}_{\rm det}$ from
eq.~(\ref{eq:23}), we need the effective monitoring time $t_{\rm eff}$ to
calculate the expected number of detections for a given
survey. If the transient emission has duration $(1+z)\Delta t_{\rm
  \infty}$ (eq.~\ref{eq:43}) and the survey has total lifespan $t_{\rm
  tot}$ and cadence $t_{\rm cad}$, the effective monitoring time is
\begin{equation}
  \label{eq:15}
  t_{\rm eff}(M, z) = \frac{t_{\rm tot}}{t_{\rm cad}}
  \mbox{min}[(1+z)\Delta t_{\infty}, t_{\rm cad}] - (1+z) \Delta t_{\infty}.
\end{equation}
Note that in eq.~(\ref{eq:15}) we have assumed: (1) the time interval
between any two consecutive exposures is always $t_{\rm cad}$; (2)
detection(s) of the transient emission must be preceded and followed by
non-detections, i.e. only the cases with ``off-on-off'' are considered
as positive signals but cases with ``on-off'' or ``off-on'' are
discarded in order to be
conservative\footnote{Since the switch-on time of the transient emission
  ($\sim r_{\rm S}/c$) is short, even if the duration $(1+z) \Delta
  t_{\infty}$ is much longer than the survey lifespan, stellar
  disruption events are still detectable as ``off-on''
  sources. They may be distinguished from other long-duration
  transients due to the smooth lightcurve and blackbody spectrum. If
  ``off-on'' sources are included, LSST may improve the limit on
  $\eta-1$ to about $10^{-8}$.}. 
Therefore, for a certain hard surface radius $\eta =
r_0/r_{\rm S}$, the expected number of detections is
\begin{equation}
  \label{eq:19}
  \begin{split}
        N_{\rm det} =&\ \int_{M_{\rm min}}^{M_{\rm max}}\mt{d} M \dot{N}(M) \\
  &\times\int_0^{z_{\rm lim}(M, \eta)}\mt{d}z \Psi(M, z)  t_{\rm eff}(M, z)
  \frac{\mt{d}V}{\mt{d}\Omega 
    \mt{d} z} \Delta \Omega. 
  \end{split}
\end{equation}

\begin{figure}
  \centering
\includegraphics[width = 0.5 \textwidth,
  height=0.25\textheight]{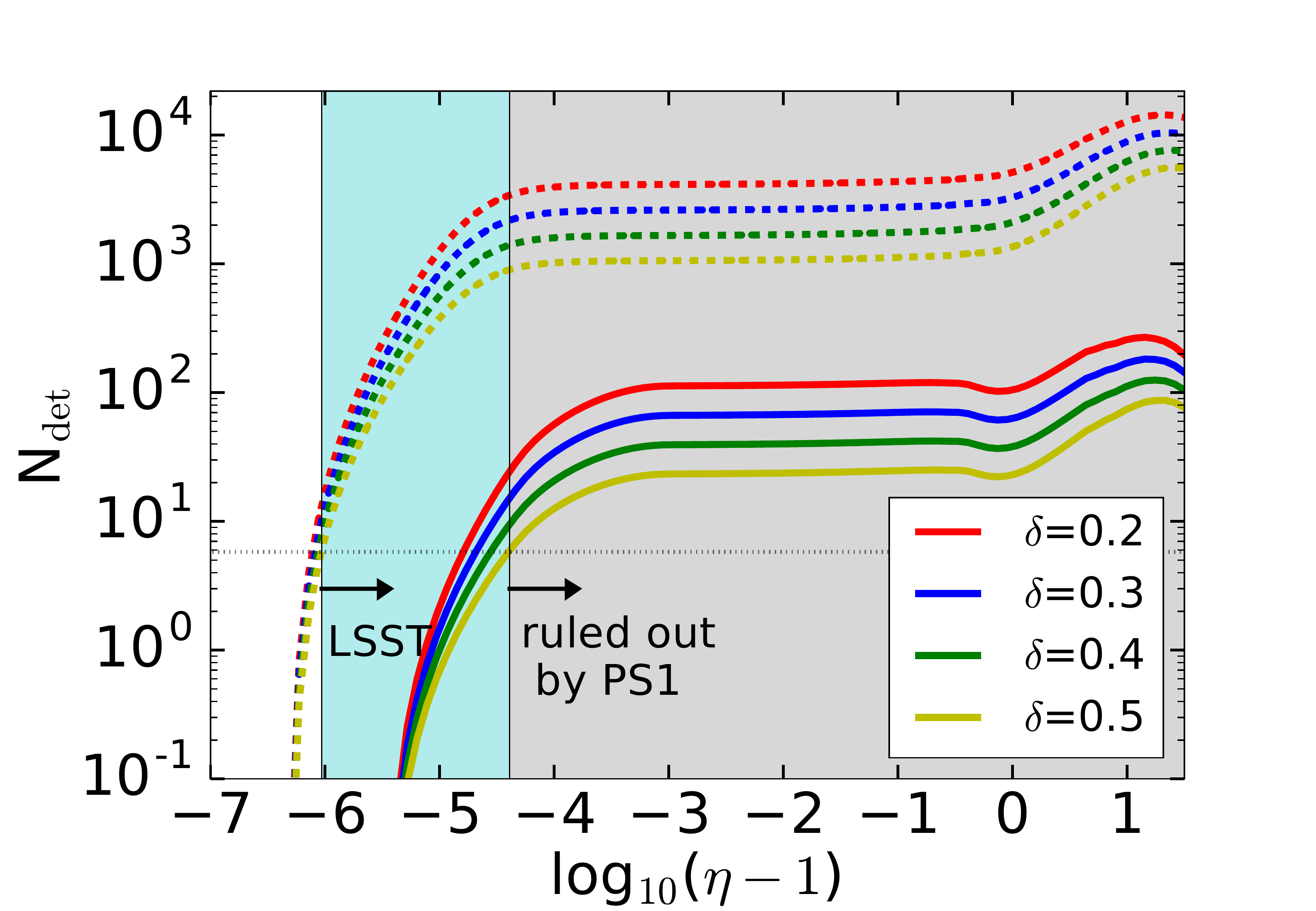}
\caption{Expected number of stellar disruption events detectable by
  Pan-STARRS1 (PS1, solid lines) and LSST (dashed lines) as a function of the
  hard surface radius $r_0 = \eta r_{\rm S}$. (Note that this plot
considers $r_0$ along the abscissa, whereas Fig.~\ref{fig:emission} 
considers the photospheric radius $r_{\rm ph}$.) Different colors
  represent different disruption rate power-law slopes ($\delta$ in
  eq.~\ref{eq:22} varying from 0.2 to 0.5). Non-detection throughout
  the survey lifespan rules out the region above $N_{\rm det}=5.81$
  (thin horizontal dotted line) at 99.7\% confidence level. For
  the conservative disruption rate power-law index $\delta = 0.5$,
  observations by PS1 have ruled out the grey shaded region $\eta -
  1>10^{-4.4}$. Future observations by LSST will be
  able to improve the limit to $\sim10^{-6}$.
  The sharp drop at the smallest $\eta$
  is caused by the duration of transient emission approaching the
  survey lifespan. The flat part in the middle is when $r_{\rm
  ph}/r_{\rm S} - 1$ approaches order of unity and we simply take $r_{\rm
  ph}/r_{\rm S} -1 = 0.3$ as a conservative limit in eq.~(\ref{eq:14})
  (larger $r_{\rm ph}$ gives higher g-band flux density). The rising
  part at $\eta- 1\gtrsim1$ is caused by the
  radiation temperature decreasing with $r_{\rm ph}$ when $r_{\rm
    ph}/r_{\rm S}-1 \gtrsim 1$ (see 
  the upper pannel of Fig.~\ref{fig:emission}). The drop when
  $\eta- 1$ approaches 30 is caused by the duration of the
  transient emission being shorter than the survey cadence. We only
  consider the parameter 
  space $1<\eta<30$ in this paper (the upper limit arises from the fact that TDEs from $\gtrsim
  10^6 M_{\rm \odot}$ CMOs have been observed). 
}\label{fig:Ndet}
\end{figure}

In Fig.~(\ref{fig:Ndet}), we show the expected number of detections for two
different surveys as a function of the hard surface radius
$r_0$. Solid lines are for g-band observations by the Pan-STARRS1 3$\pi$ survey 
\citep[PS1,][]{2010SPIE.7733E..0EK, 2013ApJ...770..128I,  
2016arXiv161205560C}, and dashed lines are for future g-band observations by the
Large Synoptic Survey Telescope 3$\pi$ survey 
\citep[LSST,][]{2008arXiv0805.2366I}. Different line 
colors represent different disruption rate power-law slopes
($\delta$ in eq.~\ref{eq:22}). Both surveys cover 3/4 of the sky, but
since low Galactic-latitude regions have significant dust extinction,
we use sky area $\Delta\Omega = 2\pi$. For a single exposure, PS1 and
LSST have g-band 5-$\sigma$ flux limit of 22.0 and 23.4 in AB magnitude,
and we only consider sources 1.5 mag brighter than the 5-$\sigma$
limits for the calculation of the number of detectable events.

PS1 3$\pi$ survey has a cadence of $t_{\rm cad}\simeq$ 3 months and
total operation time of $t_{\rm tot}\simeq$ 3.5 years (so far). LSST
3$\pi$ survey will have a cadence of $t_{\rm cad}\simeq$ 3 days and total
lifespan of $t_{\rm tot}\simeq$ 10 years. The transient searching data
products (from image subtraction) of PS1 have been 
released to the public \citep{2015ATel.7153....1H,
  2016arXiv161205243F}. If CMOs have a 
hard surface, stellar disruption events produce transients that are
distinct from traditionally known ones (e.g. supernovae, ANG,
variable stars, etc.), because they have thermal spectra with
year-long smooth lightcurves. Currently, no such transients have been
reported.

The actual number of detections follows a Poisson distribution with
expectation value $N_{\rm det}$, so non-detection rules out the region
above the horizontal thin dotted line with $N_{\rm det} = 5.81$ in
Fig.~\ref{fig:Ndet} at confidence level $1-\mathrm{exp}(-5.81)
= 99.7\%$. For instance, for the conservative case $\delta=0.5$, any
hard surface above $r_0/r_{\rm S} - 1 = 10^{-4.4}$ can be ruled out by
PS1. The lower limit depends on the slope of the stellar disruption rate
and ranges from $10^{-4.8}$ to $10^{-4.4}$ when $\delta$
goes from 0.2 to 0.5. With the same argument presented in this paper,
future observations by LSST may be able to rule out
$r_0/r_{\rm S}-1 \gtrsim 10^{-6}$ (limited by the duration of
the transient emission). We also note that only the
information from g-band is used, and if we combine g-band limits with
other bands (urizy), the constraints are slightly stronger.

\section{Discussion}\label{sec:discussion}

The main conclusions of this work can be found in the abstract and Figs.
(\ref{fig:emission}) and (\ref{fig:Ndet}). We
discuss possible issues in the analysis above. 

(1) We have assumed that all CMOs have a universal $\eta
= r_0/r_{\rm S}$ (the ratio of CMO hard surface radius
$r_0$ to the event horizon radius $r_{\rm S}$). However, the existing
data does not rule out the 
possibility that a small fraction of CMO might have
$\eta-1>10^{-4.4}$. In the future LSST era, with a much more accurate 
determination of the rate of stellar disruptions by CMOs, one should
be able to place a much stronger limit on $\eta-1$ without making the
assumption that all CMOs have the same $\eta$.

(2) We have ignored the spin of CMOs, which will modify the shape of the
hard surface and spacetime above the surface (and hence the emergent
radiation from the stellar debris). Note that the scale 
height of the layer of stellar debris is a factor of a few smaller
than $r_0-r_{\rm S}$ in the non-spinning case. And as long as the spin
is relatively slow (spin parameter $a/M\lesssim0.2$), most of the
baryonic mass is within the light 
cylinder and the structure of the stellar debris layer is not strongly
affected by rotation. The emission from the photosphere is still
determined by the fraction of the diffusive flux escaping to infinity, so
observations can rule out a similar range of 
$\eta$ as shown in Fig.~(\ref{fig:Ndet}) and the conclusion will be
similar for CMOs with $a/M \lesssim0.2$. 

(3) The situation close to the photosphere in the strong-gravity
regime is complicated: (i) the radiation field is highly anisotropic;
(ii) baryons and radiation cannot be treated as a single fluid; (iii)
the system is not adiabatic due to energy flowing in from below and
out from above; (iv) there might be large-scale convective
motion or wind\footnote{
Since the initial orbit of the star is plunging (or bound)
in the Schwarzschild spacetime, only a small fraction of the stellar
mass $M_*$ may be lost in a wind and the mass loss rate $\dot{M}_{\rm 
  w}\ll M_*/t_{\rm dif, \infty}$, where $t_{\rm dif, \infty}$ is the
diffusion time of the entire baryonic layer measured at infinity
(eq. \ref{eq:30}). If the wind speed is on the order of the local
escape velocity, it can be shown that the Thomson scattering optical
depth of the wind is $\tau_{\rm w}\simeq (r_0/r_{\rm S})^{1/2}
(\dot{M}_{\rm w} t_{\rm dif, \infty}/M_*)$, which is $\ll1$ in the
strong gravity regime.}. One caveat of
this paper is that the lightcurve of the 
transient emission is likely not flat. The hydrodynamics of the
collision between the star and the hard surface can affect the
initial lightcurve on a timescale of a few times $r_{\rm S}/c$ or
possibly as large as $\mu_{\rm ph}^{-1}r_{\rm S}/c$ (when $\mu_{\rm
  ph}\ll 1$). Then, since the flux of photons escaping to infinity is
smaller than the diffusive flux arriving at the photosphere from
deeper layers, the radiation pressure at the photosphere rises with
time causing a larger escaping flux and also pushing the photosphere
to slightly larger radii. On a timescale longer than the photon
diffusion time through the entire layer (see eq. \ref{eq:30}), the
photosphere slowly shrinks and the escaping flux decreases with time
until the radiation energy content is depleted. Solving the full
radiation-hydrodynamical structure of the 
stellar debris layer from optically thick to thin regions is left for
future work. We discuss in Appendix A the validity and limitations of
the TOV equation for describing the structure of the stellar debris layer.


(4) We were unable to provide a strong constraint on $\eta$ for CMOs of
mass $< 10^{7.5} M_{\odot}$, due to the following two reasons: (i) the
emission from the layer of stellar debris on the possible hard surface
may peak in the non-observable far UV (if $r_{\rm ph}/r_{\rm
  S}-1\sim1$); (ii) main-sequence stars are tidally disrupted before
reaching close to $r_{\rm S}$. The radiation produced (e.g. by shocks
and the accretion disk) before the gas falls onto the CMO 
makes it very hard for observations to constrain the emission we have
calculated in this work. The actual accretion rate onto the CMO is
also uncertain due to the complexities of accretion disk physics.

(5) We have assumed that baryons and radiation associated with the 
stellar debris are incorporated into the CMO's pre-existing exotic material 
(the material that forms the hard surface with which the star collides) on
a timescale $t_{\rm in}$ much longer than the duration of the transient 
radiation from stellar disruption, $\Delta t_{\infty}$, given by 
eq.~(\ref{eq:43}). 
If $t_{\rm in}$ were to be less than $\sim r_{\rm S}/c$, the layer of 
stellar debris is converted to the exotic matter before baryons can 
reach hydrostatic equilibrium, and in this case very little radiation 
will escape to infinity. If $ r_{\rm S}/c\ll t_{\rm in}< \Delta
t_{\infty}$, the debris has sufficient time to reach hydrostatic 
equilibrium and its stratification is correctly described in section
\ref{sec:profile}. However, the transient radiation from this stratified debris
does not last for the full time duration $\Delta t_{\infty}$ (calculated in
section \ref{sec:emission}), but is terminated earlier ($t_{\rm in}$)
when the transformation of the debris to the exotic matter is
completed.

(6) CMOs are growing in mass $M$ and size $r_0$ due
to gas accretion over cosmic time. To avoid the formation of an event
horizon, the mass of the baryonic layer on the hard surface must not
exceed $9\mu_0 M/8$ (see Fig.~\ref{fig:p0}). Therefore, the transformation
of radiation-baryon mixture to exotic matter must occur on a 
timescale $t_{\rm in} \lesssim \mu_0 M/\dot{M}$ ($\dot{M}$ being the accretion
rate). This assumption was implicitly made by Broderick et al. when they
considered the consequences of accretion onto a possible hard surface 
in Sgr A* and M87* \cite[e.g.][]{2015ApJ...805..179B}. Furthermore, they
assumed, based on an erroneous reasoning from the short dynamical
time ($\sim r_{\rm S}/c$), that the system can be described to be in
equilibrium such that the rate of radiation energy escaping to 
infinity is equal to the rate of mass-energy falling onto the
hard surface, i.e.  $L_{\infty}\simeq\dot{M}c^2$.
However, we point out that the dynamical time being short only
means that the baryonic layer on the hard surface is in hydrostatic
equilibrium, but it does not imply a balance between the rate of
in-falling and escaping energy. The luminosity at infinity is equal to
the accretion rate when the timescale for radiation to escape from the
hard surface is shorter than the timescale over which the accretion rate 
is roughly constant. For Sgr A*, the accretion rate likely varies on 
timescales of $t_{\rm acc} < Mc^2/L_{\rm Edd}\simeq 3.8\times
10^8\rm\ yr$. If the radiation  
from the accreted gas is released at radius $r$ where $\mu(r)\approx
r/r_{\rm S} -1 \ll1$, then only a small fraction $\mu$ of the
radiation escapes and the  
rest follows a highly curved trajectory that brings it back to the hard surface.
Therefore, photons bounce on the hard surface $\sim\mu^{-1}$
times before escaping to infinity. Thus, the time it takes for photons
to escape from the hard surface is $t_{\rm esc}\sim
\mu^{-1}r_{\rm S}/c$. For Sgr A* we have $t_{\rm acc}/t_{\rm esc}\sim
10^{14} \mu (t_{\rm acc}/10^8\rm\ yr)$, and hence if
$\mu\ll 10^{-14}$ then $L_{\infty}\ll\dot{M}c^2$. It follows from this
result that for Sgr A* a hard surface at radius $r_0/r_{\rm S} - 1\ll
10^{-14}$ cannot be ruled out. Moreover, if transformation of ordinary
matter to whatever exotic matter  makes up the hard surface occurs on
a short timescale $t_{\rm in}\lesssim r_{\rm S}/c$, very little radiation will
escape from the CMO and the object would be
indistinguishable from a BH in its electromagnetic signal.
 As pointed out by \citet{2002A&A...396L..31A}, the approach 
of Broderick et al., and the work presented here, supports the 
existence of the event horizon but does not provide a firm proof; these
works do, however, severely constrain the location of the hard surface
to be extremely close to the Schwarzschild radius, with $r_0/r_{\rm
  S}-1\lesssim 10^{-4}$ for CMOs of $M>10^{7.5}M_{\rm\odot}$ in other
galaxies, and $\lesssim 10^{-14}$ for Sgr A*.

\section{acknowledgments}
We thank Edward Robinson, Emil Mottola and James Guillochon for useful
discussions. We also thank the anonymous referee for useful comments.
WL was funded by the Named Continuing
Fellowship at the University of Texas at Austin.
RN was supported in part by NSF grant AST1312651, NASA
grant TCAN NNX14AB47G, and
the Black Hole Initiative at Harvard University, 
which is supported by a grant from the John Templeton Foundation.

\appendix
\section{}

We show that matter is well coupled to radiation in the optically
thick part of the layer, but not when the optical depth drops below
$\sim10$.

Consider an electron (associated with a proton) moving through an
isotropic radiation field. The distance it travels before being forced
to change direction by Compton scattering can be estimated
by
\begin{equation}
  \label{eq:13}
  d_{\rm sc}\sim \frac{m_pc^2}{\sigma_{\rm T} \rho},
\end{equation}
where $\rho$ is the radiation energy density, given by
\begin{equation}
  \label{eq:24}
  \rho(\mu)\simeq \frac{M_*c^2}{4\pi r_{\rm S}^3\mu_0}
  \left(\frac{\mu}{\mu_0}\right)^{-2}.
\end{equation}
Putting eq.~(\ref{eq:24}) into eq.~(\ref{eq:13}), we obtain
\begin{equation}
  \label{eq:26}
  d_{\rm sc}\sim \frac{\mu r_{\rm S}}{\tau(\mu)}\sim
  \mu^{1/2}\lambda \ll \lambda, 
\end{equation}
where $\tau(\mu) = \kappa_T M_* \mu_0/(4\pi r_{\rm S}^2\mu)$ is the
optical depth above $\mu(r)$ and $\lambda\sim \mu^{1/2}r_{\rm
  S}/\tau(\mu)$ is the local Thomson mean free path. The radiation
temperature at location $\mu(r)$ is $kT(\mu)\simeq 25 
(M_*/M_{\rm \odot})^{1/4} M_8^{-3/4} \mu_{0,-7}^{1/4}\mu_{-7}^{-1/2}\rm\
keV$, so the thermal speed of protons is non-relativistic. Therefore,
baryons diffuse very slowly and are well coupled to the local
radiation field (which dominates the energy density). This coupling
breaks down as we approach the photosphere and
$\tau(\mu)$ becomes less than about 10. Furthermore, the radiation field
and the pressure tensor become highly anisotropic for $\tau \lesssim
10$, and the TOV equation no longer provides a good description of the
structure of the layer above this point.

\label{lastpage}
\end{document}